\documentclass[aps,prl
]{revtex4}
\def\frac#1#2{{\textstyle{#1\over#2}}} %puts a small fraction in a 
\def\R{\hbox{\rm I \kern-5pt R}}

\usepackage{amssymb}
\usepackage{amsmath}
\usepackage[sort&compress]{natbib}
\usepackage{graphicx}
\DeclareGraphicsExtensions{.pdf}

\begin{document}

%%Start real stuff

\title{Quantum Tagging for Tags Containing Secret Classical Data}
\author{ Adrian Kent}
\affiliation{Centre for Quantum Information and Foundations, DAMTP, University of Cambridge, 
Cambridge, U.K.}
\affiliation{Perimeter Institute for Theoretical Physics, Waterloo, Ontario, Canada}
\date{ August 2010; revised July 2011 (extended discussion; scheme unaltered)}

\begin{abstract}
Various authors have considered schemes for {\it quantum tagging}, that is, authenticating the classical location of a 
classical tagging device by sending and receiving quantum signals from suitably located distant sites,
in an environment controlled by an adversary whose quantum information processing and transmitting power
is potentially unbounded.  
All of the schemes proposed elsewhere in the literature assume that
the adversary is able to inspect the interior of the tagging device.   All of these schemes have been
shown to be breakable if the adversary has unbounded predistributed entanglement.    
We consider here the case in which the tagging device contains a finite key string shared with distant sites but
kept secret from the adversary, and show this allows the location of the tagging device to be authenticated 
securely and indefinitely.    Our protocol relies on quantum key distribution between the tagging device and
at least one distant site, and demonstrates a new practical application of quantum key distribution.
It also illustrates that the attainable security in position-based cryptography can depend crucially on 
apparently subtle details in the security scenario considered.   
\end{abstract}
\maketitle
\section{Introduction} 

There is now a great deal of theoretical and practical interest in the
possibility of basing unconditionally secure cryptographic tasks on 
some form of no signalling principle as well as, or even instead of, the laws
of non-relativistic quantum theory.   The earliest examples of which we are
aware are bit commitment protocols based on no signalling \cite{kentrel, kentrelfinite}, 
which are provably secure against all classical attacks
and against Mayers-Lo-Chau quantum attacks.   
Another significant development was the  first secure quantum key distribution
protocol based on no signalling \cite{bhk}, which was followed by other
schemes and stronger security results \cite{agm,sgbmpa,asp,mrwbc,masanes,hrw}.  
Other interesting examples include a protocol for variable bias coin tossing \cite{vbct}, 
and recent results based on the no-summoning theorem \cite{nosummoning}, including
a protocol for location-oblivious data transfer \cite{otsummoning} and 
a simple practical and provably unconditionally secure quantum relativistic protocol
for bit commitment \cite{bcsummoning}.   

The idea of {\it quantum tagging} -- authenticating the location of a tagging device (or {\it tag}) by exchanging
classical and quantum information with a number of suitably located sites -- has recently 
been independently considered by several authors\cite{taggingpatent, malaney, chandranetal, kms}. 
A scheme which offers security against current technology, but is not unconditionally secure,
was patented in 2006 \cite{taggingpatent}.   Other schemes initially presented as unconditionally
secure\cite{malaney, chandranetal} were recently proposed, but subsequently shown to be
breakable by teleportation-based attacks \cite{kms}.    These attacks were significantly
extended \cite{buhrmanetal} and shown to apply to a large class of tagging schemes, including
some discussed in Ref. \cite{kms}, where their security was left as an open question, and others
conjectured to be secure in Ref. \cite{laulo}.   
However, it appears that all the schemes proposed in Refs. \cite{taggingpatent, malaney, chandranetal, kms, laulo} 
may be secure against eavesdroppers who are unable to predistribute and store entanglement; this is
proven \cite{chandranetal, buhrmanetal} for the schemes of Ref. \cite{chandranetal}.  
All of these schemes implicitly or explicitly assume a security model in which an adversary
can send signals through the tag, and can observe its interior workings, but cannot manipulate information within it.  

Quantum tagging, and other tasks in position-based quantum cryptography \cite{chandranetal, buhrmanetal},
raise new questions about cryptographic security assumptions.
It seems reasonable to think of the remote sites which exchange signals with the tag as secure laboratories, 
to which the standard cryptographic assumptions for sending and receiving stations apply.   
In particular, it seems reasonable to assume that classical and quantum data can be generated, processed
and stored securely within these sites, in a way that prevents an adversary from reading classical data,
measuring quantum data, or interfering with any processing.    

On the other hand, there is more than one sensible and practically relevant security scenario for the tag.  
One can imagine scenarios in which the sending and receiving stations
are large and cryptographically hardened, while the tag is designed to be small, is operating in a 
more hostile environment, and therefore cannot utilize the same resources or guarantee the same 
level of security.    It may make sense in such a scenario to assume that an adversary can send
signals through a tag, and inspect its interior at will, even if the tag is protected well enough that direct
physical sabotage is not possible (at least within a short enough time interval 
to allow the tagging protocol to be spoofed).   
Equally, though, one can imagine scenarios in which the tagging device and 
sending and receiving stations are all functionally equivalent -- for instance, are all satellites --
and should all be covered by the same security model.    In this case, the standard cryptographic
assumptions imply that {\it both} the tag and the distant sites can keep data secret from the adversary. 

Practical considerations aside, quantum tagging raises interesting theoretical questions 
when considered as a more abstract problem about the possible query-response tasks that an
agent can carry out given control of only a subset $R$ of Minkowski space. 
In this context too it is interesting to consider different types of tasks, in which, for 
example, the agent may or may not be allowed to send signals through the complement $\bar{R}$
of $R$ and infer information about localized physical states in $\bar{R}$.  

All of these points suggest it is worth thinking carefully about the various possible security models
for quantum tagging.   An even stronger motivation comes from recent results of \cite{buhrmanetal} 
showing that a large class
of quantum tagging schemes are not unconditionally secure, and their conjecture that these
results extend to all quantum tagging schemes within the security model considered to date.
It would clearly be advantageous to have unconditionally secure tagging schemes if they are available within a reasonable security model.    It would also be very interesting, for both theoretical and practical reasons, to 
understand what are the minimal security resources required to ensure unconditionally secure
tagging. 

In this paper we show that unconditionally secure quantum tagging is indeed possible and can be sustained
indefinitely within a sensible
security scenario, in which the tag, like the sending and
receiving stations, is able to keep classical data secret from the adversary. 
We give a simple secure tagging protocol in this scenario, which uses quantum key distribution
(more precisely, quantum key expansion) between one (or more) site(s) and the tag, and 
then uses the generated secret key as a resource for secure authentication of the tag's
position.    The protocol defines a new, simple and practical application of quantum key
distribution.     Our protocol appears reasonably efficient, but we do not know whether
it is close to optimally efficient, or whether perhaps significantly greater efficiency can
be achieved by other means, for example by protocols that do not use quantum key
distribution as an explicit sub-protocol.    This is an interesting and potentially practically
significant open question.   

\section{Previously considered scenarios} 

We work within Minkowski space-time, $M^{(n,1)}$, with $n$ space dimensions.
The most practically relevant case is $n=3$.\footnote{Corrections due to 
general relativity in weak gravitational fields can be made without qualitatively 
affecting our discussion\cite{kentrelfinite}.  In particular, our discussion
applies to tagging protocols on or around Earth, assuming that it is reasonable
to neglect exotic gravitational phenomena 
such as wormholes or attacks based on major distortions of the local space-time geometry.}
However, we will initially consider the case of one space dimension, which simplifies the discussion while
illustrating all the key points.   Tagging schemes rely on using signalling constraints to 
give a lower bound on the distance between the tag and remote sites. 
In our scheme, once one sees this can be done securely for a tag between two 
remote sites in one dimension, it is easy to see that the same applies for a tag
within the convex hull of $n+1$ remote sites in $n$ dimensions, using multilateration. 

The following two security scenarios for quantum tagging were defined in Ref. \cite{kms},
from which (to avoid any possible confusion by altering language) we quote verbatim.   
We  take these as the standard definitions for the task as analysed to date: earlier discussions in
the academic literature \cite{malaney,chandranetal} were less than fully explicit about
their security scenario, and subsequent work \cite{buhrmanetal, laulo} to date has followed
these definitions, implicitly or explicitly.   

\subsection{Security scenario I}

Alice operates cryptographically secure sending and receiving
stations $A_0$ and $A_1$, located in small
regions (whose size we will assume here is negligible, to simplify
the discussion) around distinct points $a_0$ and $a_1$ on the real line. 
The locations of these stations are known to and trusted by Alice,
and the stations contain synchronized clocks trusted by Alice.
Her tagging device $T$ occupies a finite region $ [ t_0 , t_1 ]$
of the line in between these stations, so that $a_0 < t_0 < t_1 < a_1 $.   The tagging device
contains trusted classical and/or quantum receivers, computers and
transmitters, which are located in a small region (which again, to
simplify the discussion, we assume is 
of negligible size compared with $(t_1 - t_0)$ and other parameters)
around the fixed point $t_+ = \frac{1}{2} ( t_0 + t_1 )$.
The device is designed to follow a protocol in which classical
and/or quantum outputs are generated via the computer from inputs defined
by the received signals.   The outputs are sent in one or both directions, left
and/or right (i.e. towards $a_0$ and/or $a_1$), this choice again in general 
depending on the inputs.  The tagging device may also contain a trusted clock,
in which case the clock time is another allowed input.

We assume that signals can be sent from $A_i$ to $T$, and within $T$, at
light speed, and that the time for information processing within $T$ 
(or elsewhere) is negligible.   $T$ is assumed {\it immobile} and {\it physically secure}, in the sense that an
adversary Eve can neither move it nor alter its interior structure.
However, $T$ is not assumed {\it impenetrable} by Eve: Eve may be able to 
send signals through it at light speed.  Nor (for now -- we will reconsider this
below) is it assumed {\it cryptographically
secure}: Eve may also be able to inspect its interior.   In particular, 
$T$ contains no classical or quantum
data which Alice can safely assume secret, and she must thus assume
that its protocol for generating outputs from inputs is potentially
public knowledge.

$T$ can be switched on or off.   When switched off, it remains immobile
and physically secure, and simply allows any signals
sent towards it to propagate unmodified through it: in particular, signals
travelling at light speed outside $T$ also travel through $T$ at light speed. 

Eve may control any region of space outside $A_i$ and $T$, may send classical
or quantum signals at light speed through $A_i$ and $T$ without $A$ (or $T$) detecting them, may
be able to jam any signals sent by $A_i$ or $T$, and may carry out arbitrary 
classical and quantum operations, with negligible computing time, anywhere
in the regions she
controls.   Eve cannot cause any information processing to take place within $T$,
other than the (computationally trivial) operation of transmitting arbitrary signals through
$T$, except for the operations that $T$ is designed to carry out on appropriate input signals.   
Her task is to find a strategy which {\it spoofs} the actions of $T$, that is, 
makes it appear to $A$ that $T$ is switched on when it is in fact
switched off.    Conversely, $A$'s task is to design $T$, together with a tagging
protocol with security parameter $N$, so that the chance, $p(N)$, of 
$E$ successfully spoofing $T$ throughout a given time interval $\Delta t$
obeys $p(N) \rightarrow 0$ as $N \rightarrow \infty$.   

\subsection{Security scenario II}

In scenario II, the tag is physically secure, but not immobile.  
Eve can move it, without disturbing its inner workings, at any
speed up to some bound $v$, known to Alice.   
Clearly $v=c$, the speed of light, gives an absolute upper
bound.   To avoid considering relativistic effects, 
we assume $v \ll c$ in this scenario.

\section{New security scenario: a cryptographically secure tag}

While the above scenarios may be the most reasonable in some important contexts,
they are not the only interesting ones.    As we have already noted above,
 in quantum key distribution schemes, and other cryptographic schemes
based on physics, one normally assumes that Alice's laboratories are {\it cryptographically
secure}.  That is, Eve cannot to gain any information about the physics taking place within these 
laboratories or affect it in any way: she cannot read any classical or quantum data generated within the 
laboratories, nor alter any of the devices within the laboratories that generate or process data.   

To be sure, whether or not this is justified in any given context is a significant empirical question. 
However, cryptography is useless if one cannot generate or process information
securely anywhere.    And if it is reasonable to suppose that the remote sites $A_i$ can be
made cryptographically secure, it is also reasonable (in appropriate scenarios) to suppose that
$T$ can too.     So, we now consider variants of the above security models,
in which both the $A_i$ and $T$ are cryptographically secure.   Specifically, and crucially,
we suppose that $T$ can contain a finite secret classical bit string, preshared with and 
hence known to $A$, which $E$ is unable to read by any means.   

Note that we do {\it not} suppose that $T$ is {\it impenetrable} to Eve: we suppose that she 
may be able to send signals at light speed through $T$ (and indeed, though it is less relevant,
through the $A_i$).     There are at least two good reasons for this.   First, impenetrability would be a
logically stronger assumption, which (it turns out) we do not need to make, and we
are interested in identifying the minimal necessary assumptions here.   
Second, the distinction is sensible in practice: one can easily imagine regions secure enough
that Eve cannot, practically speaking, get a usefully detailed image of their interior, but which are 
nonetheless semi-transparent to signals on some frequency or of some type.
In fact, every putatively secure cryptographic site has the latter feature (as a reductio, 
consider the possibility of Eve signalling with gamma ray bursts, neutrino beams, or high energy shock waves).  

\subsection{A secure tagging protocol using a cryptographically secure tag}

We need one further assumption: that $A$ and $T$ were at one point able to share a finite secret random 
bit string while maintaining its secrecy.  This seems reasonable for many, perhaps most, applications of tagging:
if Alice has $T$ within one of her laboratories at some point prior to the tagging protocol, she can include the shared random 
key at this point.  Given this, and given that the $A_i$ and $T$ are cryptographically secure, $A$ and $T$ can carry out a
secure quantum key distribution -- or, more precisely, quantum key expansion -- protocol, using the initial shared random
key to authenticate their communications and thus guarantee the iterative generation of longer shared random keys. 
This process can continue alongside, and throughout, any tagging protocol.   We can thus assume that, although $T$ need
only have finite data storage capacity, $A$ and $T$ can effectively generate over the course of time an 
arbitrarily long shared secret random bit key $k = k_0 k_1 k_2 \ldots$. 

Consider the following tagging scheme (recall that for now we are working in
one space dimension). 

1. $A$ sends a series of independently randomly chosen bits, $a_i$, from $A_0$, and 
another series of independently randomly chosen bits, $b_i$, from $A_1$.
These signals are sent at light speed, timed so as to arrive pairwise
simultaneously at $t_+$: that is, $a_1$ and $b_1$ arrive together, then $a_2$ and $b_2$, 
and so on. 

2. On receiving $a_i$ and $b_i$, $T$ immediately retrieves the key bit $k_{4i + 2 a_i + b_i}$ from memory storage,
 and sends this bit at light speed to
both $A_0$ and $A_1$.  

3. $A$ tests that the outputs are correct and arrived at the appropriate times.
If this test is passed for $N$ successive bit pairs, sent within the interval $\Delta t$, 
she accepts the location of $T$ as authenticated. 

This scheme is evidently secure under both scenarios.   In scenario I, if $T$ is switched off, then $T$ will not 
release the relevant key bits (even if $T$ was able to generate them before being switched off). 
$E$ thus has no information about which key bit value to broadcast, and will send the wrong value
with probability $1/2$ per query.   

In scenario II, the relevant key bit is known only to $A$ and $T$ until $T$ 
receives instructions to broadcast it.  $T$ will only ever release and 
broadcast one out of the four bits $k_{4i}, k_{4i +1}, k_{4i +2}, k_{4i + 3}$. 
$E$ cannot identify which of these bits should be broadcast until she knows both $a_i$ and $b_i$; if she generates
and supplies her own input on one side (or both) to a relocated $T$ before the correct bit arrives from $A$, she risks obtaining an incorrect
 key bit from the string as output, in which case $T$ will never 
release the correct key bit.
Any relocation of $T$ thus either introduces a detectable delay in the responses
to $A_0$ or $A_1$, or (with probability $1/2$ per query) causes incorrect key bit values to be sent, or both.

\begin{figure}[t]
\centering
\includegraphics[width=120mm]{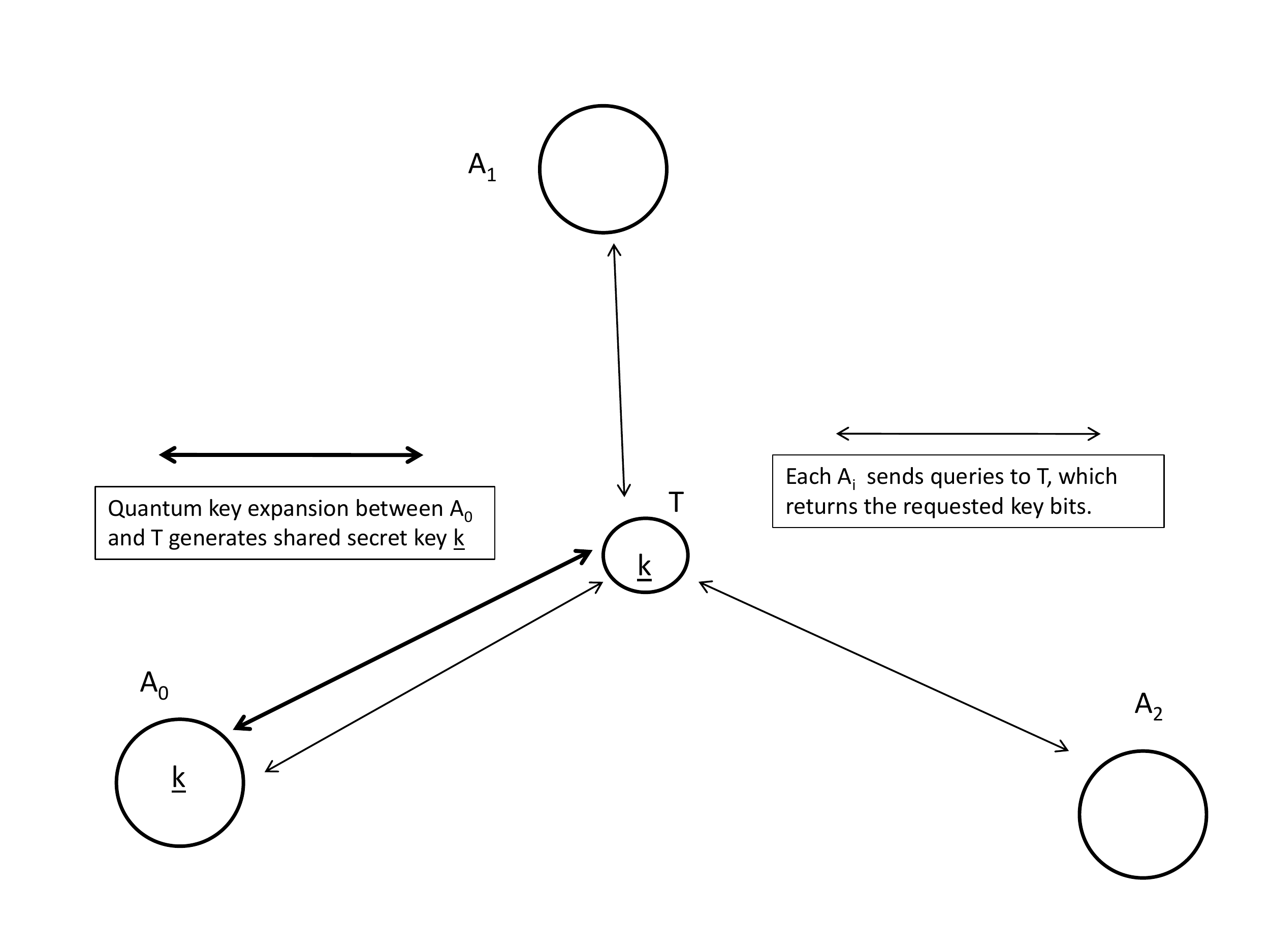}
\caption{One implementation of secure tagging in two dimensions.  Here
  the key is generated by quantum key expansion between $A_0$ and $T$.
  $A_0$ shares the key with $A_1$ and $A_2$ either via secure
  communication based on quantum key expansion, or by transmitting
  relevant key bits after they have been queried.}
\end{figure}

\section{Comments}

It might at first sight seem puzzling that secure tagging requires a
protocol involving anything more than secure communication channels, since one
might think that the tag can identify its own position using something
like GPS technology and then securely broadcast this information to
Alice over shared secure channels.   Note, however, that in all our
scenarios, we make the cryptographically standard assumption that
everything except $T$ and $A$'s laboratories is potentially under the
control of Eve, and hence cannot be trusted.  In particular, $T$
cannot trust any remote GPS system to give accurate readings.   
More generally, we cannot (without giving $T$ more internal resources) assume that
$T$ initially knows its own position.  

We have seen that tagging can nonetheless be securely implemented using a quantum protocol --
essentially bootstrapping on the security of quantum key expansion -- under the assumption of a
cryptographically secure tag.   This scheme has the advantage that it does not require 
$T$ to contain even a trusted clock.   It can easily be extended to higher dimensions, using
suitably located sending and receiving stations to authenticate each of
the tag's coordinates.   An efficient way to do this is by authenticated multilateration
within the convex region spanned by Alice's laboratories. 

For example, in three dimensions, Alice could proceed as follows.
She sets up four laboratories $A_i$ which do not
lie in a common plane.  Using quantum key expansion, she generates a shared secret key
with the tag, which is split into four separate keys $k_i$, which themselves are split
into length two blocks.   

This does not necessarily require QKE to be implemented
between all four $A_i$ and the tag: a QKE link from $A_0$ to $T$ suffices.   $A_0$ can
share the keys $k_i$ with the other $A_i$ using QKE links from $A_0$ to $A_i$.  
Alternatively, bits from these keys can be shared over an authenticated, but not necessarily
secret, link, after $A_i$ has (allegedly) received the relevant bits from $T$.   This 
adds to the delay in authenticating position, but does not compromise its security.

Each $A_i$ sends at light speed requests to $T$ for randomly chosen bits from successive blocks of the key $k_i$,
and $T$ immediately returns the requested bits.   This allows each $A_i$ to obtain an authenticated bound
on its distance from $T$.   Comparing these bounds allows the $A_i$ to authenticate the location
of $T$, provided that $T$ lies within the tetrahedron that they span.    

Clearly, the scheme could be modified in many ways. 
One interesting option is to use the shared key
generated by quantum key expansion to authenticate the tagging messages sent between
$A$ and $T$.  While this is not necessary for the security of the scheme, it does 
make security even more transparent, since authenticating all transmissions effectively eliminates
the possibility of successful spoofing.   It also has the advantage of making it much harder for Eve (or
random noise) to interfere with the tagging protocol by sending fake inputs to the tag --
although, of course, if Eve has sufficiently advanced technology, resources and determination
she can jam any quantum cryptography scheme, and in particular any tagging scheme.  
It might too have some efficiency advantages in some scenarios. 

Our protocol is intrinsically quantum, in the sense that it requires quantum information to be 
sent in at least one direction between at least one of the $A_i$ and $T$, in order to use the
power of quantum key expansion to generate an arbitrarily long shared secret bit string from
the initial finite shared secret bit string.   This allows the tagging scheme to continue indefinitely.
Without the use of quantum key expansion, the duration of the tagging scheme is limited by 
the size of the secure memory that the tag can contain.   Of course, a finite duration classical tagging
scheme may well be useful in some scenarios, just as a finite classical one-time pad can be useful
for sending secret messages of limited length.    In both cases, the power of using quantum information
is that it allows secure transmissions to be continued indefinitely, even though the physically secure 
resources initially (and indeed at any later time) are necessarily finite.

\section{Acknowledgements}

This work was partially supported by an FQXi mini-grant and by Perimeter Institute for Theoretical
Physics. Research at Perimeter Institute is supported by the Government of Canada through Industry Canada and
by the Province of Ontario through the Ministry of Research and Innovation.
I thank William Munro, Jonathan Oppenheim, Timothy Spiller and Damian Pitalua-Garcia for helpful conversations.  
%
%refs
%

\end{document}